\documentclass[reprint,prl,amsmath,amssymb,aps,longbibliography,lengthcheck,showpacs]{revtex4-1}
\newcommand{\beq}{\begin{equation}}
\newcommand{\eeq}{\end{equation}}
\newcommand{\bea}{\begin{eqnarray}}
\newcommand{\eea}{\end{eqnarray}}
\newcommand{\uvec}[1]{{\bf \hat{#1}}}
\newcommand{\eq}[1]{{Eq. (\ref{#1})}}

\usepackage{graphicx}
\usepackage{dcolumn}
\usepackage[caption=false]{subfig}
\usepackage{bm}
\begin{document}
%
\title{Knot soliton solutions for the one-dimensional non-linear Schr\"{o}dinger equation}
\author{Rahul O. R.}\email{rahulor.14@iist.ac.in} \author{S. Murugesh}\email{murugesh@iist.ac.in}
%
%
\affiliation{Department of Physics, Indian Institute of Space Science
	and Technology, Thiruvananthapuram-695~547, India.}
\date{\today}
%
\begin{abstract}We identify that for a broad range of parameters a variant of the soliton solution of the one-dimensional non-linear Schr\"{odinger} equation, the {\it breather}, is distinct when one studies the associated space curve (or soliton surface), which in this case is knotted.  The significance of these solutions with such a hidden non-trivial topological element is pre-eminent on two counts: it is a one-dimensional model, and the nonlinear Schr\"{o}dinger equation is well known as a model for a variety of physical systems.  	

\pacs{05.45.Yv,75.10.Hk,75.30.Ds}
\end{abstract}
\maketitle
\section{Introduction}
\label{intro}
The one-dimensional nonlinear Schr\"{o}dinger equation (NLSE) is a fundamental model naturally and frequently arising in a 
variety of physical systems such as fluid dynamics, dynamics of
polymeric
fluids, ferromagnetic spin chains, fiber optics and vortex dynamics in superfluids, to name a few 
\cite{hasi:1972,hopfinger:1982,vinen:2008,bird:1987,hase:1989,schw:1988,ml:1977}. Besides its physical importance, it also is a very important
model in soliton theory, owing to its rich mathematical structure. It is 
completely integrable with soliton solutions, and presents itself amenable to nearly
every method available in the study of nonlinear systems, making it a perfect
pedagogical model. Its complete integrability was first established in
a classic paper by Zakharov and Shabat in 1972, which also brought
about a deeper clarity, from a geometric point of view, on the method of 
inverse scattering transform being developed around that period\cite{zs:1972,sulem:1999}. For these aforementioned reasons it remains one of the most studied, and among
the most understood of nonlinear integrable systems. 
Inspite of its rich history and continued interest, investigations on NLSE have often thrown out novel results and physical behavior never anticipated or intuited earlier. One such is the {\it breather} solition, a solitonic behavior that shows up periodically in time or space, often referred to as Ma and Akhmediev breathers, respectively\cite{ma:1979,akhm:1986}. It also displays a {\it rogue}
behavior as a special case 
marked by a sudden, momentary yet colossal enhancement in 
the field amplitude\cite{pere:1983}. 
Interest in this special solution have grown manifold since citing of
similar
phenomenon in the deep ocean\cite{kharif:2008}. As a point of further interest,  the same have
been
predicted to occur in Bose-Einstein condensates, whose dynamics is
modeled by the Gross-Pitaevskii equation, which closely resembles the
NLSE in higher dimension\cite{akhm:2009}. 

Furthermore, a natural
connection exists between the complex field described by NLSE in one dimension and a space curve in 3-d, whose time evolution spans a surface. 
In the case of systems integrable and endowed with a
Lax pair the {\it soliton surfaces} thus obtained are of much
interest  in soliton theory\cite{hasi:1972,ml:1977,sym:1985,schief:2002}.  Often the space curves thus obtained are by themselves  
physically realizable, adding to their
significance\cite{sm:2005}. In particular, the space curves for the NLSE field can be thought of as thin filament vortices in fluids, or superfluids\cite{hasi:1972, hopfinger:1982,vinen:2008}. This association leads to a connected question  whether the  one dimensional NLSE can possibly possess a soliton solution that can be related to a curve with non-trivial topology, such as  one with a knot. In fact, Kleckner and Irvine experimentally showed that such knotted vortices can indeed be created in fluids\cite{kleck:2013}, which further raises curiosity concerning  the existence of such solutions to the one-dimensional NLSE. 
  
The space curves associated with the breather solitons have been obtained explicitly by Cie\'{s}li\'{n}ski et. al. in \cite{sym:1986}. While the breather is qualitatively a periodic (temporally, spatially, or both---the rogue being a breather with both the temporal and spatial period tending to infinity) version of a {\it usual} soliton, the corresponding space curve is a closed loop, carrying in it a smaller traveling loop that folds around the larger loop as it travels. But, to our knowledge, no solutions of the NLSE have been reported thus far that is associated with a knotted curve.

A variant of the Akhmediev breather solitons, owing to a certain Galilean gauge admissibility of the NLSE, have  been studied by various authors\cite{dysthe:1999,sal:2013}. Further, the associated space curve has been constructed by numerically integrating the evolution equations for the Frenet-Serret triad of orthonormal vectors by Salman\cite{sal:2013}.  However, a more detailed investigation of these breather solitons, particularly an explicit expression for these curves {\it in their full generality},  proves more worthwhile. As a case in point we show that for a wide range of parameters, these soliton solutions are characteristically different in that the associated space curve is {\it knotted} --- a simple {\it overhand} knot that occurs periodically (as is expected of a breather). 
Although localized stable knots have been encountered in the study of complex nonlinear systems, especially NLSE { like} systems (see for instance \cite{niemi:1997}), what separates the solitons presented here is that these exact solutions are associated with the {\it one-dimensional} NLSE, wherein structures with a non-trivial topology are not expected.

\section{Knotted breathers for the NLSE}
\label{sec:2}
The one-dimensional NLSE for a complex field  $\psi(x,t)$ is given by
\beq\label{nlse}
i\psi_t + \psi_{xx} + 2|\psi|^2\psi =0, 
\eeq
wherein the subscripts $t$ and $x$ indicate derivatives with respect to {\it time} and {\it space}, respectively. The variable $\psi(x,t)$ could refer to the complex amplitude of the electric field, in the context of light transmitted through an optical fiber with quadratic nonlinearity\cite{hase:1989}, or a complex function of the curvature and torsion in the case of thin vortices in a fluid, etc\cite{sm:2005}. $\psi_0=0$ is clearly a trivial solution for the NLSE, \eq{nlse}. Starting from this {\it seed} solution, one 
may proceed with any of the standard techniques
of obtaining solitons, the direct method due to Hirota, or a Darboux transformation, to obtain 
a soliton solution --- a localized traveling wave of the secant hyperbolic type\cite{atlee:1992,schief:2002}. Instead, if we start with another simple seed solution of the NLSE,
\beq\label{mpa}
\psi_0 = \kappa_0 e^{2i\kappa_0^2t},
\eeq 
for some constant $\kappa_0$, one may derive a periodic breather solution\cite{ma:1979,pere:1983,akhm:1986}. 

In this paper we investigate in detail the soliton solutions obtained if one starts with  another non-trivial seed solution
\beq\label{seed}
\psi_0=\kappa_0 e^{i\sqrt{2}\kappa_0x}
\eeq
 ---a constant field of uniform magnitude with a spatially periodic phase. 
Proceeding by any of the standard techniques, one can obtain, after some detailed algebra, the three parameter one-soliton solution 
\beq\label{1sol}
\psi_1 =  e^{i \sqrt{2} \kappa_0 x} \Big(\kappa_0 - 2\, \lambda_{0I}\, \frac{(\zeta - i\, \eta )}{\chi} \, \Big)
\eeq
wherein
\bea\label{1sol_a}
\zeta   = c_1 \, \cos(2\, \Omega_{0R}) + c_2 \, \cosh(2\, \Omega_{0I}) \nonumber\\
\eta    = c_3 \, \sin(2\, \Omega_{0R}) - c_4 \, \sinh(2\, \Omega_{0I}) \nonumber\\
\chi   = c_2 \, \cos(2\, \Omega_{0R}) + c_1 \, \cosh(2\, \Omega_{0I}) \nonumber\\
\Omega_0  =\Omega_{0R} + i\Omega_{0I}     = f_0 \, ( x - \sqrt{2} \mu_0 t )\nonumber\\
f_0   = f_{0R} + i f_{0I}         = \frac{1}{\sqrt{2}} \, \sqrt{\nu_0^2 + 2\, \kappa_0^2}\nonumber\\
\mu_0 = \mu_{0R} + i\mu_{0I} = \kappa_0 - \sqrt{2}\, \lambda_0 \nonumber\\
\nu_0 = \nu_{0R} + i\nu_{0I} = \kappa_0 + \sqrt{2}\, \lambda_0 \nonumber\\
c_1 = 2 \, \big( 4\, \kappa_0^2 + 2\, |\nu_0|^2 + 4 \sqrt{2}\, \kappa_0 \, \nu_{0I} + 4\, |f_0|^2 \big)\nonumber\\
c_2 = 2 \, \big( 4\, \kappa_0^2 + 2\, |\nu_0|^2 + 4 \sqrt{2}\, \kappa_0 \, \nu_{0I} - 4\, |f_0|^2 \big)\nonumber\\
c_3 = 2 \, \big( 8\, \kappa_0 \, f_{0I} + 4\sqrt{2}\, (\nu_{0R} \, f_{0R} + \nu_{0I} \, f_{0I})\,\big)\nonumber\\
c_4 = - 2 \, \big( 8\, \kappa_0 \, f_{0R} + 4\sqrt{2}\, (\nu_{0I} \, f_{0R} - \nu_{0R} \, f_{0I})\,\big)\nonumber\\
\eea

and $\lambda_0 =\lambda_{0R}+i\lambda_{0I}$ is an arbitrary complex spectral parameter associated with 
the one-soliton (the scattering parameter in the framework of inverse scattering transforms). Two more parameters indicating the initial position and phase of the soliton are taken to be zero, without any loss of generality. 

 It may be noted that the seed solution for the breather, \eq{mpa}, differs from the seed we have chosen in \eq{seed} only by a phase factor. Under a Galelian transformation to the NLSE, the field parameter $\psi$ gets phase shifted. More specifically, the NLSE is invariant under  the transformation 
 \bea\label{gali}
 x\to x-vt,\,\,t\to t,\nonumber\\
 \psi\to\psi e^{i(vx/2+v^2t/4)}.
 \eea
 Thus, starting from the spatially periodic Akhmediev breather, one may directly obtain a Galelian gauge related counterpart simply by effecting such a transformation\cite{dysthe:1999,sal:2013}.  Indeed, for the choice 
 $\lambda_{0R}=-\kappa_0/\sqrt{2}$ and $\kappa_0^2 > \lambda_{0I}^2$, the general solution we have presented in \eq{1sol_a} reduces to such a breather obtained by
 Salman \cite{sal:2013}. Alternately, when $f_{0I}\mu_{0R} + f_{0R}\mu_{0I}=0$ it reduces to the Galilean transformed version of the temporally periodic Ma breather. Being a complex function in one dimension, the profile of the breather does not fully reveal its intricacies. The  profile, in fact, is qualitatively similar to the Akmadiev or Ma breathers. But, as pointed out earlier, the complex field $\psi$ can be inherently and systematically related to a curve in three dimensional euclidean space. Such a curve can display non-trivial topology, and in this case forms an overhand knot in the process of its time evolution.

\section{Knotted Breather Space Curves}
It is well known that
the complex field of the NLSE, $\psi(x,t)$, can be linked systematically to a moving { non-stretching} curve in three dimensions. Thus, one way of describing such a curve through NLSE is to relate $\psi$ to the  intrinsic curvature $\kappa$ and torsion $\tau$ of the curve\cite{eisen:1909}, such as\beq\label{kt}
\psi = \frac{\kappa}{2} e^{i\sigma},\,\, \sigma_x=\tau,
\eeq
where, importantly, $x$ now represents the arc-length parameter of the curve. 
Often referred as the Hasimoto transformation, this form arises quite naturally in certain cases --- for instance, in studying the motion of a thin vortex filament in a fluid\cite{hasi:1972}. If ${\bf R}(x,t)$ were such a space-curve, the NLSE, \eq{nlse}, can be rewritten in terms of ${\bf R}$ as 
\beq\label{lia}
{\bf R}_t = {\bf R}_x\times{\bf R}_{xx},
\eeq 
--- the {\it localized induction approximation} (LIA)\cite{saff:1992}. It should be noted that, whereas in the NLSE $x$ represented the spatial coordinate, in the LIA though the same is the arc-length parameter of a non-stretching curve. Consequently, while a Galilean transformation to the NLSE may be countered by an appropriate phase transformation to $\psi$ (see \eq{gali}), it is not as straight forward in the case of the LIA. For instance, the transformation \eq{gali} amounts to changing the torsion of the curve by a constant factor while retaining its curvature ($\kappa\to\kappa$, $\tau\to \tau+v/2$). In the language of the curve vector ${\bf R}(x,t)$ this is non-trivial, and certainly not achieved by the co-ordinate transformation given in \eq{gali}.

The fundamental theorem of curves guarantees the existence of a unique curve, given $\kappa$ and $\tau$ (upto a global shift, or rotation). While the general solution for such a curve for a given $\kappa$ and $\tau$ is  unknown, for special cases though, such as when the complex function $\psi$ is a soliton solition obtainable by inverse scattering, the associated curve and surface can indeed be systematically found\cite{sym:1985}. 
To a reasonable approximation, thin line vortices in incompressible fluids and superfluids are often interpreted as such curves associated with the NLSE \cite{hopfinger:1982,vinen:2008}. 
The curves associated with the Ma and Akhmadiev breathers have been investigated by Cie\'{s}li\'{n}ski, et. al
in\cite{sym:1986}, while that for the Galilean transformed Akhmediev breather has been numerically studied in \cite{sal:2013}. 

The seed solution in \eq{seed} corresponds to a 
moving helix with a constant intrinsic curvature $\kappa_0$ and a torsion $\tau = \sqrt{2}\kappa_0$:
\bea\label{helix}
	{\bf R}_0 =
	\frac{1}{3} \bigg[\bigg(  \sqrt{3}( x + 2 \sqrt{2} \kappa_0 t ) \bigg) \uvec{i}
	+\bigg(\frac{1}{\kappa_0} \sin\theta )\bigg)                                         \uvec{j}\nonumber\\
	-\bigg( \frac{1}{\kappa_0} \cos\theta \bigg)\uvec{k}\bigg],\,\,\,\,\,\,\,\,
	\eea
where $\theta = \sqrt{6}\kappa_0(x-\sqrt{2}\kappa_0 t)$, having  both a global translation along its axis with velocity $\frac{2\sqrt{2}}{3}\kappa_0$, and a rotation about its axis with period
\beq\label{t_0} 
T_0=\pi/(\sqrt{3}\kappa_0^2),
\eeq
effectively constituting a screw motion. The helix has a pitch $\sqrt{2}\pi/(3\kappa_0)$ and radius $1/(3\kappa_0)$. 

For the one-soliton solution in \eq{1sol}, the associated curve can be found to be
\begin{widetext}
\begin{equation}
\label{curve}
	{\bf R}_1= {\bf R}_0  +\frac{\lambda_{0I}}{|\lambda_0|^2\chi}\bigg[-\frac{(\sqrt{2}\eta+\xi)}{\sqrt{3}}
	\uvec{i}
	+\bigg(-\zeta\sin\theta +\cos\theta\frac{(\eta-\sqrt{2}\xi)}{\sqrt{3}}\bigg)\uvec{j}
		+\bigg(\zeta\cos\theta +\sin\theta\frac{(\eta-\sqrt{2}\xi)}{\sqrt{3}}\bigg)\uvec{k}\bigg]
\end{equation}
\end{widetext}
where 
\beq
\xi = c_4 \, \sin(2\, \Omega_{0R}) + c_3 \, \sinh(2\, \Omega_{0I}),
\eeq
and 
$\zeta$, $\eta$, $\chi$, and the constants $c_i,\,i=1-4$ were defined in \eq{1sol_a}. They also obey the 
conditions
\bea\label{cons}
\zeta^2 + \eta^2 + \xi^2 = \chi^2,\nonumber\\
c_2^2 +c_3^2+c_4^2 =  c_1^2. 
\eea
What makes this soliton solution, \eq{1sol}, really distinct is that, for a range of values, the filament self intersects, developing into an {\it overhand} knot, and back into an un-knotted loop, as it evolves in time (Fig. 1).  The overall period of the breather, $T_{total}$, is thus divided into two phases --- the {\it knot} phase with period $T_{knot}$, and the {\it loop} phase with period $T_{loop}$. While $\kappa_0$ determines both the radius and pitch  of
the helical backbone (see \eq{helix}), the radius of the soliton loop is determined by both $\kappa_0$ and $\lambda_{0I}$. When the loop is larger than the pitch of the backbone, folding results in self intersections with the helical backbone, with the outcome being periodic knot formation.   In general, this behavior is multiply periodic, both temporally and spatially, involving periods that are generally incommensurate. To see this, we first note that the helical back bone in itself has a periodicity $T_0$ decided by $\kappa_0$, \eq{t_0}.  Besides, due to the conditions in 
\eq{cons}, the terms in the curve equation, \eq{curve}, can be re-written as:
\bea
\frac{\sqrt{2}\eta+\xi}{\sqrt{3}\chi} &\equiv \sin A\cos B, \nonumber \\
\frac{\eta-\sqrt{2}\xi}{\sqrt{3}\chi} &\equiv \sin A\sin B, \nonumber \\
\frac{\zeta}{\chi} &\equiv \cos A, 
\eea
for appropriate real quantities $A$ and $B$, determined by $\kappa_0$ and $\lambda_0$. 
Thus ${\bf R}_1$ is a function of three generally incommensurate periodic terms. 
However, for carefully chosen values of 
$\kappa_0$ and $\lambda_{0}$, the periodicity could be exclusively temporal, or spatial. 
\begin{figure*}[]
	\subfloat[$t=t_1$]%
	{\includegraphics[clip,trim=0.3cm .7cm .5cm 2.2cm,width=0.32\textwidth]{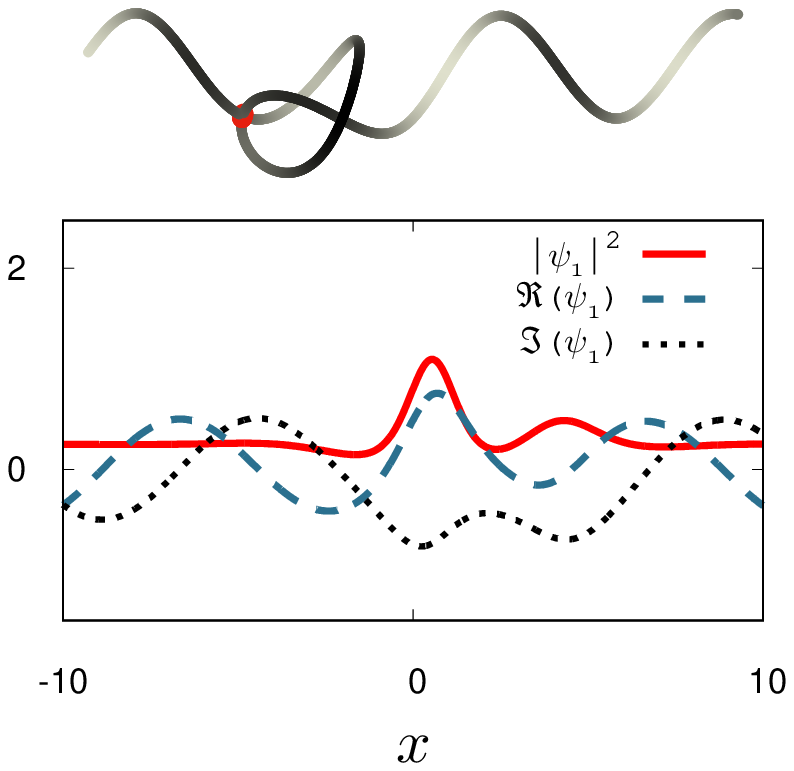}}
	\hskip .35cm
	\subfloat[The {\it loop} phase]%
	{\includegraphics[clip,trim=0.3cm .7cm .5cm 2.2cm,width=0.32\textwidth]{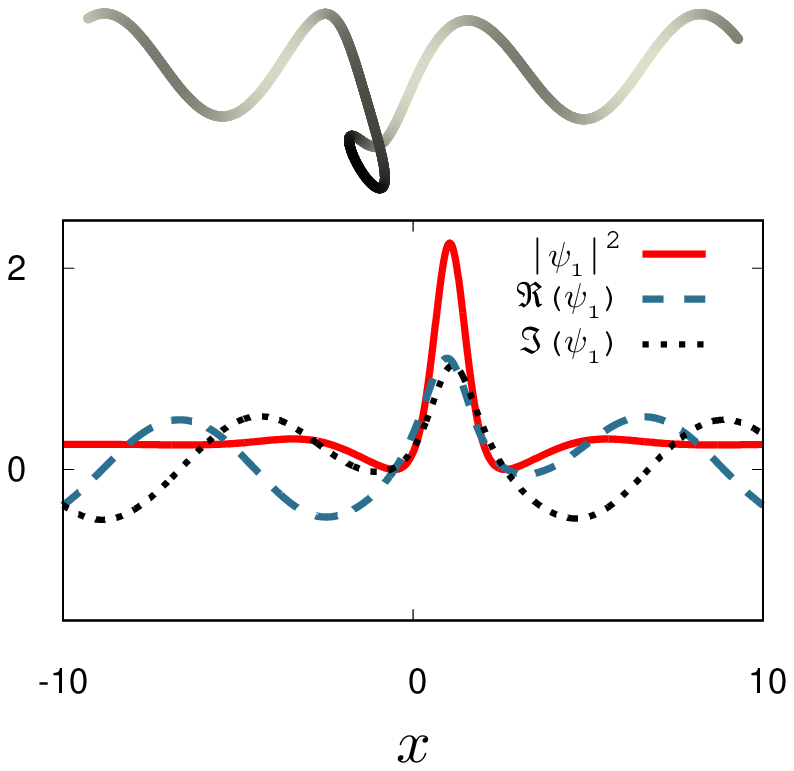}}
	\hskip .35cm
	\subfloat[$t=t_2=t_1+T_{loop}$]%
	{\includegraphics[clip,trim=0.3cm .7cm .5cm 2.2cm,width=0.32\textwidth]{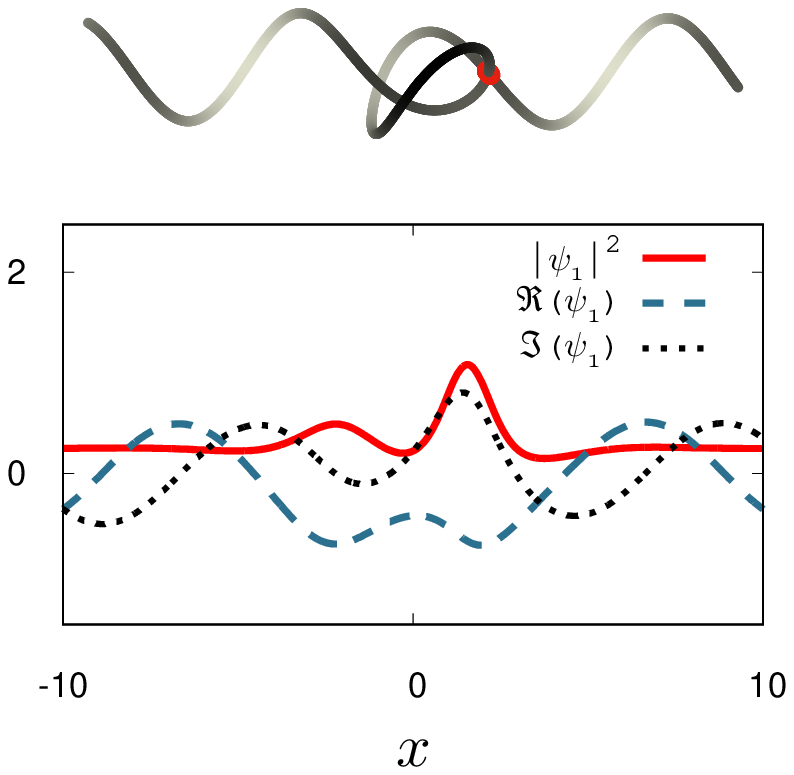}}	\\
		\subfloat[The {\it knot} phase]%
		{\includegraphics[clip,trim=0.3cm .7cm .5cm 2.2cm,width=0.32\textwidth]{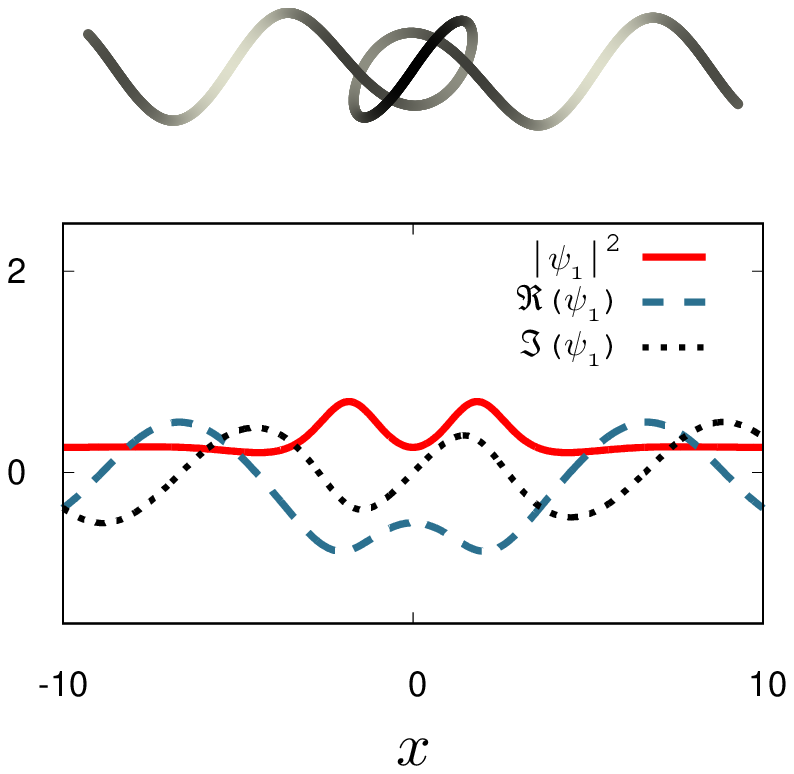}}
		\hskip .35cm
		\subfloat[$ t = t_2 + T_{knot} = t_1 + T_{total}$]%
		{\includegraphics[clip,trim=0.3cm .7cm .5cm 2.2cm,width=0.32\textwidth]{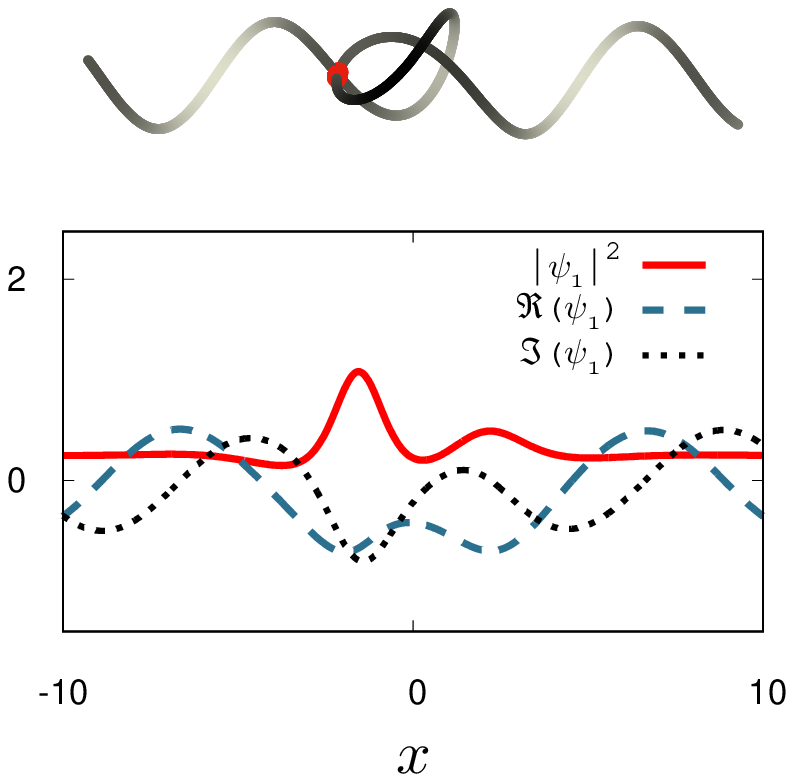}}
		\hskip .35cm
		\subfloat[$T_{total}=T_{loop}$,\, $T_{knot}=0$]%
		{\includegraphics[clip,trim=0.3cm .7cm .5cm 2.2cm,width=0.32\textwidth]{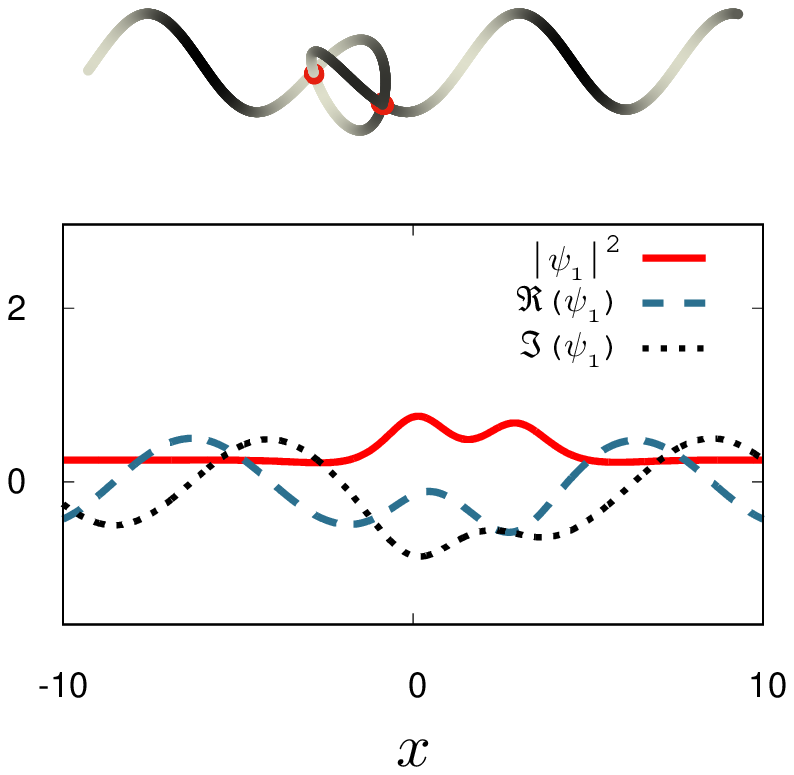}}
\caption{(a)-(e): The space curves associated with the 1-soliton solution \eq{1sol}, and the corresponding energy ($|\psi|^2$(x)) profile, for $\kappa_0=0.5$ and $\lambda_{0I}=0.5$  through an entire period of its evolution, $T_{total}$. The real and imaginary parts of $\psi$ are shown by dashed and dotted lines, respectively. The period is marked by three successive self intersections (red dot on the curves in (a), (c) and (e)), subdividing it into two phases --- the {\it knot} phase with period $T_{knot}$, and the loop phase with period $T_{loop}$. For this choice of parameters, the filament forms a loop for the period between the first two intersections, and a knot for the remaining period. Energy localization is more pronounced in the loop phase than in the knot phase. The curve in (a) is same as (e), but for a global rotation due to the period $T_{total}$ being incommensurate with the rotation period of the helical backbone.  (f) Two simultaneous intersections (shown here for $\lambda_{0I}=0.6$), with a vanishing $T_{knot}$. See Supplementary Materials \cite{knot} and \cite{two_int} for detailed animation on (a)-(e) and (f).}	
\end{figure*}

For a specific choice of the parameters $\kappa_0$ and $\lambda_{0I}$ (for ease of analysis, henceforth we shall choose  $\lambda_{0R}=0$, and freeze the value of $\kappa_0$, varying only $\lambda_{0I}$), snapshots of this curve through its period are shown in Fig. 1 (a)-(e) --- a loop traveling along a helical backbone of the seed, folding around the axis of the helix as it travels(see Supplementary Material \cite{knot} for detailed animation). In the process, it also encounters self intersections, leading to formation of a overhand knot. The behavior is not generic, however. For certain choices of $\lambda_{0I}$, given $\kappa_0$, the two self intersections are simultaneous at two different points on the curve, with a vanishing $T_{knot}$ (see Fig. 1 (f), and supplementary Material \cite{two_int} for detailed animation). 

The dependence of $T_{knot}$ and $T_{loop}$ on $\lambda_{0I}$ is cumbersome to be analyzed analytically from \eq{curve}. In Fig. 2 we show this dependence following a numerical study.
Generally, $T_{loop}>T_{knot}$. The overall period of  evolution $T_{total}$ is better elucidated  if one transforms to a new varying arc-length parameter 
\beq\label{trans}
x' = x - \frac{\sqrt{2}}{f_{0I}}(\mu_{0R}f_{0I}+f_{0R}\mu_{0I})t,
\eeq 
and choosing $x'=0$. 
Recalling the role of $x$ in LIA as the arc-length parameter, this transformation amounts to moving along the curve at a constant speed such that $\Omega_{0I}=0$. Consequently, $\zeta,\,\eta,\,\xi$ and $\chi$ in \eq{curve} are periodic functions of time whose period 
\beq
T_{total} = 
\frac{\pi f_{0I}}{\sqrt{2}\mu_{0I}|f_{0}|^2} = 
\frac{\pi f_{0I}}{2\lambda_{0I}|f_{0}|^2}
\eeq
determines the total period of the loop evolution. 
Besides, there are bands of $\lambda_{0I}$ values for which no self intersections are noted. Beyond a certain value of $\lambda_{0I}$, intersections lead only to loops. Heuristically, this can 
be attributed to the size of the loop being smaller than the pitch of the helical backbone  for large values of $\lambda_{0I}$, as a simple dimensional argument would indicate. 
\begin{figure}[t]
	\centering
	\scalebox{0.9}{\input{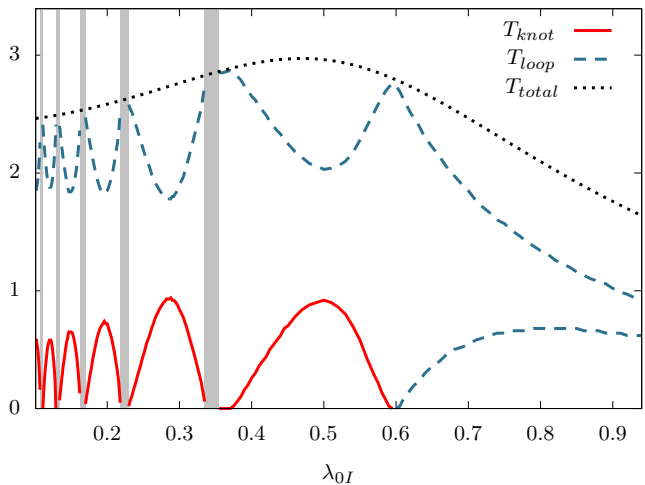}}
	\caption{Time period of loop evolution as a function of $\lambda_{0I}$, while $\kappa_0$ is held constant (at 0.5). A period $T_{total}$ (dotted line) is marked by three successive self intersections (see Fig. 1). In the interval between two successive intersections the filament may remain a loop, or turn into a knot, with respective time periods $T_{loop}$ (dashed lines)  $T_{knot}$ (solid lines). $T_{total}=T_{loop}+T_{knot}$.  No intersections are noticed for certain intermediate ranges of $\lambda_{0I}$, indicated by gray bands (see Supplementary Material \cite{no_int} for detailed animation). There are points where $T_{knot}$ vanishes, corresponding to two simultaneous self intersections.  No knots are formed for values beyond $\lambda_{0I}=0.6$, although intersections do occur. }
\end{figure}
Being solutions of the LIA, it is indeed tempting to associate these knotted filaments with possible configurations of thin fluid vortices. We caution though that in order for the transition from a loop phase to the knot phase and back to happen the filament has to seamlessly go through self intersections. This at best reveals a limitation of the LIA as a model for inviscid fluid vortices. The LIA is obtained as a approximate
expression for the velocity of a thin vortex filament when one starts from the more general Biot-Savart law inspired expression\cite{saff:1992}
\beq\label{biot}
{\bf R}_t(x) \propto \int \frac{d{\bf R}(x')\times({\bf R}(x)-{\bf R}(x'))}{|{\bf R}(x)-{\bf R}(x')|^3}dx'.
\eeq 
A self intersection is {\it not} forbidden, {\it per se}, in LIA, as it simply ignores any long distance interaction. In a real fluid, such an intersection will lead to a bifurcation of the filament at the intersection point into a closed loop and a linear filament vortex. In Kleckner and Irvine's  experimental demonstration of trefoil knots and linked ring vortices the life time is indeed limited by the occurrence of intersections, which breaks them into separate loops. It should be feasible, following \cite{kleck:2013}, to generate a helical vortex filament with an overhand knot, as in Fig. 1. The life time of the knot is controlled by the parameters $\kappa_0$ and $\lambda_{0I}$ (see Fig. 2), which in practical terms can be tuned by the pitch and radius of the helical backbone, and the size of the knot. Its life time would culminate with a self intersection, when the knotted filament is expected to split into a helical filament and a vortex ring.   The more complete Biot-savart expression for the filament velocity, \eq{biot}, also dictates a similar bifurcation, as revealed numerically in \cite{sal:2013}. Besides fluid vortices, being a fundamental model for a variety of physical systems (optical propagation in nonlinear media, or 1-d ferromagnetic chains, for instance) it is fair to surmise that these knotted breathers for the NLSE, howsoever short lived,  bear a wider scope and significance.  


\end{document}